# Actively tuning anisotropic light-matter interaction in biaxial hyperbolic material $\alpha$-MoO$_3$ using phase change material VO$_2$ and graphene

Kun Zhou[1,2], Yang Hu[3,4], Xiaoxing Zhong[1,2,*] and Xiaohu Wu[3,**]

[1]Key Laboratory of Gas and Fire Control for Coal Mines (China University of Mining and Technology), Ministry of Education, Xuzhou, 221116, China

[2]School of Safety Engineering, China University of Mining and Technology, Xuzhou, 221116, China

[3]Shandong Institute of Advanced Technology, Jinan, 250100, China

[4]Basic Research Center, School of Power and Energy, Northwestern Polytechnical University, Xi'an, Shanxi 710072, China

*Corresponding author: zhxxcumt@163.com

**Corresponding author: xiaohu.wu@iat.cn

**Abstract**

Anisotropic hyperbolic phonon polaritons (PhPs) in natural biaxial hyperbolic material $\alpha$-MoO$_3$ has opened up new avenues for mid-infrared nanophotonics, while active tunability of $\alpha$-MoO$_3$ PhPs is still an urgent problem needing to be solved. In this study, we present a theoretical demonstration of actively tuning $\alpha$-MoO$_3$ PhPs using phase change material VO$_2$ and graphene. It is observed that $\alpha$-MoO$_3$ PhPs are greatly dependent on the propagation plane angle of PhPs. The metal-to-insulator phase transition of VO$_2$ has a significant effect on the hybridization PhPs of the $\alpha$-MoO$_3$/VO$_2$ structure and allows to obtain an actively tunable $\alpha$-MoO$_3$ PhPs, which is especially obvious when the propagation plane angle of PhPs is 90°. Moreover, when graphene surface plasmon sources are placed at the top or bottom of $\alpha$-MoO$_3$ in $\alpha$-MoO$_3$/VO$_2$ structure, tunable coupled hyperbolic plasmon–phonon polaritons inside its Reststrahlen bands (RBs) and surface plasmon–phonon polaritons outside its RBs can be achieved. In addition, the above-mentioned $\alpha$-MoO$_3$-based structures also lead to actively tunable anisotropic spontaneous emission (SE) enhancement. This study may be beneficial for the realization of active tunability of both PhPs and SE of $\alpha$-MoO$_3$, and facilitate a deeper understanding of the mechanisms of anisotropic light-matter interaction in $\alpha$-MoO$_3$ using functional materials.

## I. Introduction

Hyperbolic materials (HMs) exhibit unique optical, electronic and magnetic properties, its natural hyperbolicity offers advantages over other conventional metal and dielectric materials [1-6]. As a special class of HMs, biaxial $\alpha$-phase molybdenum trioxide ($\alpha$-MoO$_3$) has drawn much attention due to its enigmatic optoelectronic characteristics throughout the visible and infrared regions [7-9]. $\alpha$-MoO$_3$ composing of octahedral unit cells with nonequivalent Mo-O bonds along three principal crystalline axes (between 550 cm$^{-1}$ and 1000 cm$^{-1}$) has recently attracted huge attention owing to its ability to excite highly anisotropic phonon polaritons (PhPs) with in-plane hyperbolic and elliptic dispersion [10], which gives rise to a versatile platform for studying anisotropic light-matter interaction. Because of its biaxial anisotropy and hyperbolicity, $\alpha$-MoO$_3$ is an excellent candidate to provide more rich physical phenomena than other HMs, thus an in-depth understanding of $\alpha$-MoO$_3$ is greatly significant for many different fields involving light-matter interactions, such as near-field radiative heat transfer, super-resolution technologies, spontaneous emission (SE) and spectral absorption [11-14]. However, the active tunability of PhPs excited by resonance between photons and $\alpha$-MoO$_3$ is still a challenge owing to their inherent crystal lattice resulting in inefficient tunability.

Stacking and combining functional materials into a heterostructure offers a flexible and promising way to couple material properties and reveal underlying physical phenomena [15, 16]. As a typical functional material, because of its electrostatic doping possibility and ability to generate higher confinement and lower losses [15, 16], graphene becomes an excellent candidate for the realization of tunable surface plasmon polaritons (SPPs) via applying extra gate voltage in the mid-infrared and terahertz frequency ranges [17-19]. Meanwhile, graphene can be flexibly combined with other micro/nanostructures and HMs to obtain coupled hybridization characteristics, which provides an alternative route to realize tunable optoelectronic devices [20, 21]. Over the past few years, graphene has been applied to realize tunable hyperbolic plasmon–phonon dispersion, tunable near-field radiative heat transfer, SE enhancement and spectral absorption [22-26]. Except for graphene, phase-change materials (PCMs) including Ge$_2$Sb$_2$Te$_5$ (GST) and vanadium dioxide (VO$_2$) have witnessed abundant achievements toward actively tunable light-matter interaction from the perspectives of principles and applications [15, 16]. Compared to that from amorphous GST to crystalline GST at about 438 K [16], the phase-transition temperature of VO$_2$ from insulator to metal is about 340 K [27], which is easy to achieve and has little influence on the corresponding cooperation materials. Meanwhile, the phase-transition temperature of VO$_2$ can be tuned by doping VO$_2$ flake with W, Ti, Mg, and Al nanoparticles forming M$_x$V$_{1-x}$O$_2$ (M = W, Ti, Mg or Al) [28-32], such as the transition temperature can be tailored to 295 K by setting the composition $x$ at 1.5% in W$_x$V$_{1-x}$O$_2$ [32]. Because its advantages over other PCMs, VO$_2$ serves as the basis for potential applications such as temperature-adaptive radiative coating, energy storage and reconfigurable electronics [33, 34].

Since the dispersion modes including phonons and plasmons can be tuned in layered van der Waals material systems using functional materials [15, 16], in this study, a theoretical platform will be constructed to actively tune anisotropic PhPs in biaxial HM $\alpha$-MoO$_3$ using VO$_2$ and graphene. Hybridization PhPs dispersion modes of α-MoO$_3$ and coupling hybridization dispersion between $\alpha$-MoO$_3$ PhPs and graphene SPPs based on phase change material VO$_2$ substrate will be detaily discussed, meanwhile the propagation plane angle in biaxial $\alpha$-MoO$_3$ will also be considered. Furthermore, comparative analysis of actively tunable anisotropic SE enhancement in $\alpha$-MoO$_3$-based structures will also be investigated.

## II. The geometry and numerical model

In order to actively tune and control PhPs behavior of the biaxial HM α-MoO$_3$ along its main crystallographic directions, VO$_2$ layer is used as the substrate owing to its attractive dielectric properties during the temperature-tuned insulator-to-metal phase transition [28]. As a proof of concept, the above-mentioned schematic is displayed in Fig. 1(a), where a Cartesian coordinate system is established, $x$, $y$ and $z$ axes respectively represent the [100], [010] and [001] crystal directions of the $\alpha$-MoO$_3$; $\phi$ is the propagation plane angle of PhPs with respect to the [100] crystal direction, $t$ is defined as the $\alpha$-MoO$_3$ thickness. $\alpha$-MoO$_3$ is a natural biaxial HM and its optical response is dominated by the phonon absorption, whose principal permittivity components can be expressed using following Lorentz equation [9]

$$\varepsilon_j = \varepsilon_\infty^j \left( 1 + \frac{{\omega_{LO}^j}^2 - {\omega_{TO}^j}^2}{{\omega_{TO}^j}^2 - \omega^2 - i\omega\Gamma^j} \right), \qquad (1)$$

where $j = x, y, z$ denotes the three crystal principal axes, corresponding to the crystalline directions [100], [001] and [010] of α-MoO₃, respectively. $\varepsilon_\infty^j$ represents the high-frequency dielectric constant, $\omega_{LO}^j$ and $\omega_{TO}^j$ refer to the longitudinal and transverse optical phonon frequencies, $\Gamma^j$ is the broadening factor of the Lorentzian line shape, the corresponding parameters can be referred from Ref. [9].

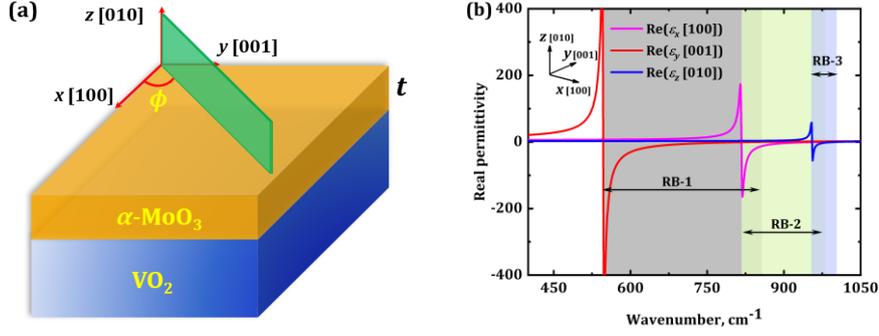

Fig. 1. (a) Schematic of illustration of the α-MoO₃/VO₂ structure. (b) Real parts of the principal permittivity tensor components of α-MoO₃.

The real parts of the dielectric tensor components $\varepsilon_j$ in α-MoO₃ are displayed in Fig. 1(b), three Reststrahlen bands (RBs) exist in the mid-infrared range of 545 to 1010 cm⁻¹. RB-1, RB-2 and RB-3 of α-MoO₃ are respectively in the ranges of 545 to 851 cm⁻¹ (the gray region), 820 to 972 cm⁻¹ (the green region) and 958 to 1010 cm⁻¹ (the yellow region), which originate from in-plane phonon mode along the [001], [100] and [010] crystalline directions, respectively. It is worth noting that RB-1 and RB-2 belong to the type-II hyperbolic band, while RB-3 is the type-I hyperbolic band. In addition, the relative permittivity tensor of α-MoO₃ for the propagation angle $\phi$ is obtained from $\varepsilon(\phi) = R(\phi)\varepsilon_{\alpha-MoO_3}R^T(\phi)$, where $R(\phi)$ is the rotation matrix for transformation of coordinate system, and the relative permittivity tensor of α-MoO₃ can be written with [35, 36]

$$\varepsilon(\phi) = \begin{pmatrix} \varepsilon_x \cos^2\phi + \varepsilon_y \sin^2\phi & \sin\phi\cos\phi(\varepsilon_y - \varepsilon_x) & 0 \\ \sin\phi\cos\phi(\varepsilon_y - \varepsilon_x) & \varepsilon_x \sin^2\phi + \varepsilon_y \cos^2\phi & 0 \\ 0 & 0 & \varepsilon_z \end{pmatrix}. \quad (2)$$

Since graphene supports SPPs in the mid-infrared region, graphene layers are added at the top or bottom of α-MoO₃ layer to realize additional PhPs tunability of the α-MoO₃ and investigate coupling hybridization dispersion between α-MoO₃ PhPs and graphene SPPs. In this study, graphene is treated as an ultrathin dielectric layer with the thickness $\Delta$ = 0.34 nm by an effective permittivity $\varepsilon = 1 + i\sigma_g/(\omega\varepsilon_0\Delta)$ [24], $\varepsilon_0$ is vacuum permittivity, and its conductivity $\sigma_g$ modeled as a sum of the intraband and interband terms is given by $\sigma_g = \sigma_{intra} + \sigma_{inter}$ being written with [24]

$$\sigma_{intra} = \frac{i}{\omega + \frac{i}{\tau_g}} \frac{2e^2 k_B T}{\pi \hbar^2} \ln\left[2\cos\left(\frac{\mu_g}{2k_B T}\right)\right], \quad (3)$$

$$\sigma_{inter} = \frac{e^2}{4\hbar}\left[G\left(\frac{\hbar\omega}{2}\right) + \frac{4i\hbar\omega}{\pi}\int_0^\infty \frac{G(\xi) - G\left(\frac{\hbar\omega}{2}\right)}{(\hbar\omega)^2 - 4\xi^2} d\xi\right]. \quad (4)$$

here $G(\xi) = \sinh(\xi/k_B T)/[\cosh(\mu_g/k_B T) + \cosh(\xi/k_B T)]$. $\tau_g$, e, $k_B$, $\mu_g$, $\hbar$ and $T$ are the relaxation scattering time, electron charge, Boltzmann constant, chemical potential, reduced Planck's constant, and temperature, respectively.

### III. Results and discussion
### 3.1 Actively tuning biaxial PhPs of *α*-MoO₃ based on VO₂ substrate

The dispersion relations of the above-mentioned heterostructures are calculated by an intensity of the imaginary part of the Fresnel reflection coefficient $Im(r_{total}^{pp})$ [15, 16], where $r_{total}^{pp}$ is the Fresnel reflection coefficient for *p*-polarized wave, which is calculated using the 4×4 matrix [37-39]. As shown in Fig. 2, the hybridization PhPs dispersion of the α-MoO₃/VO₂ structure along its main crystallographic directions at the metal- and insulator-state of VO₂ are calculated. Meanwhile, since α-MoO₃ is a biaxial HM, the effects of the propagation plane angle $\phi$ on the PhPs dispersion modes of α-MoO₃ are also considered at $\phi$ = 0°, $\phi$ = 45° and $\phi$ = 90°, respectively. The thickness of the α-MoO₃ layer is 150 nm to ignore the quantum confinement effects [9]. It is worth noting that the color bars of the hybridization dispersion modes are appropriately adjusted to clearly show the differences. When VO₂ is in the metal-state shown in Figs. 2(a)-2(c), two hybridization PhPs dispersion modes of the α-MoO₃/VO₂(I) structure at $\phi$ = 0° are gathering inside RB-2 and RB-3, and the dispersion modes lead to group velocity $v_g$ = dw/dg>0 [16] in RB-2 and $v_g$<0 in RB-3, respectively. Two PhPs dispersion modes of the α-MoO₃/VO₂(I) structure at $\phi$ = 90° occur at RB-1 and RB-3, the group velocity in RB-2 keeps the same sign with that at $\phi$ = 0°, meanwhile the group velocity is $v_g$>0 in RB-1. Interestingly, the PhPs dispersion modes of the *α*-MoO₃/VO₂(I) structure at $\phi$ = 45° within its hyperbolic bands simultaneously occur, while these PhPs dispersion modes just span the part frequency range of its RBs, which can be attributed to the overall consideration of the rotated permittivity tensor between the [100] and [001] crystalline directions [35, 36]. Similar results can also obtain as VO₂ are in the insulator-state shown in Figs. 2(d)-2(f), while the hybridization PhPs dispersion modes of *α*-MoO₃ are different before and after VO₂ phase transition. Regardless of the phase transition state of VO₂, the hybridization PhPs dispersion modes in *α*-MoO₃ exhibit obvious in-plane anisotropic behaviors.

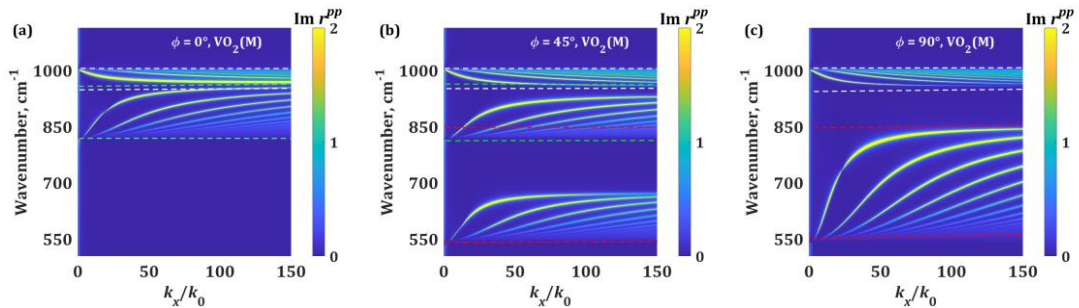

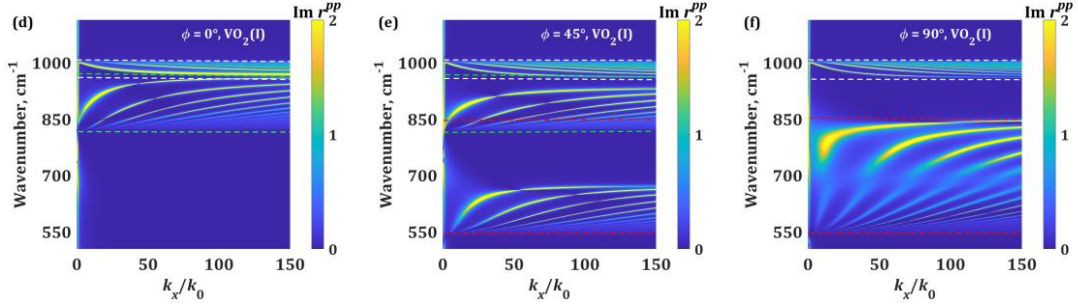

Fig. 2. Hybridization dispersion of the α-MoO$_3$/VO$_2$ structure for different propagation plane angles $\phi$ at (a)-(c) metal- and (d)-(f) insulator-state of VO$_2$.

As an example, hybridization PhPs dispersion modes of the α-MoO$_3$/VO$_2$(M) and α-MoO$_3$/VO$_2$(I) structures at $\phi = 90°$ are analyzed, as shown in Fig. 2(c) and Fig. 2(f). Since the metal-to-insulator transition has a profound impact on the dielectric properties of VO$_2$, compared to that in the α-MoO$_3$/VO$_2$(M) structure, the hybridization PhPs branches in the α-MoO$_3$/VO$_2$(I) structure significantly approach to the light line and broaden, which can be attributed to its inherent lossy enhancement of VO$_2$ once it undergoes a phase transition from the metal-to-insulator state. Furthermore, in order to clearly describe that VO$_2$ can actively tune α-MoO$_3$ PhPs, hybridization PhPs dispersion of the α-MoO$_3$/VO$_2$(M) and α-MoO$_3$/VO$_2$(I) structures varying with the wavevector components $k_x$ and $k_y$ at the wavenumber of 780 cm$^{-1}$ are calculated in Figs. 3(a)-3(b). The excitation range of hybridization PhPs dispersion branches are within $-\sqrt{-\varepsilon_x/\varepsilon_y} < k_y/k_x < \sqrt{-\varepsilon_x/\varepsilon_y}$ [40], the corresponding dielectric function components of α-MoO$_3$ at the wavenumber of 780 cm$^{-1}$ are $\varepsilon_x = 20.7 - i0.8$ and $\varepsilon_y = -1.95 - i0.07$. As shown in Figs. 3(a) and 3(b), the bright color within the regions up and down the origin being bounded by the two white dashed lines of $k_y \approx \pm 3.26 k_x$ is the hybridization PhPs dispersion branches. Once VO$_2$ undergoes a metal-to-insulator phase transition at the wavenumber of 780 cm$^{-1}$, the complex dielectric function of VO$_2$ will convert from $\varepsilon_{VO_2(M)} = -10.91 + i103.48$ to $\varepsilon_{VO_2(I)} = 0.72 + i0.96$, thus the dielectric function differences of VO$_2$ between metal- and insulator-state leading to the hybridization PhPs dispersion modes in the α-MoO$_3$/VO$_2$ structure are different, and its hybridization PhPs dispersion branches gradually move towards to the directions of a smaller vector. Furthermore, the analytical expression for the dispersion relations of the α-MoO$_3$/VO$_2$ structure under the quasistatic approximation is utilized and can be written with [12]

$$\frac{k_\rho}{k_0} = \frac{\xi}{k_0 t}\left[\arctan\left(\frac{\varepsilon_1 \xi}{\varepsilon_z}\right) + \arctan\left(\frac{\varepsilon_{VO_2} \xi}{\varepsilon_z}\right) + l\pi\right], \ l = 0,1,2\cdots \quad (5)$$

here $\varepsilon_1$ and $\varepsilon_{VO_2}$ are the permittivity of air and VO$_2$, $t$ is the α-MoO$_3$ flake thickness, $\varepsilon_z$ is the α-MoO$_3$ permittivity along the [010] crystalline directions, $\xi = i\sqrt{\varepsilon_z/(\varepsilon_x \cos^2\phi + \varepsilon_y \sin^2\phi)}$. As shown in Figs. 3(a)-3(b), the hybridization PhPs dispersion branches of the α-MoO$_3$/VO$_2$ structure with green dashed lines agree well with that using the 4×4 matrix.

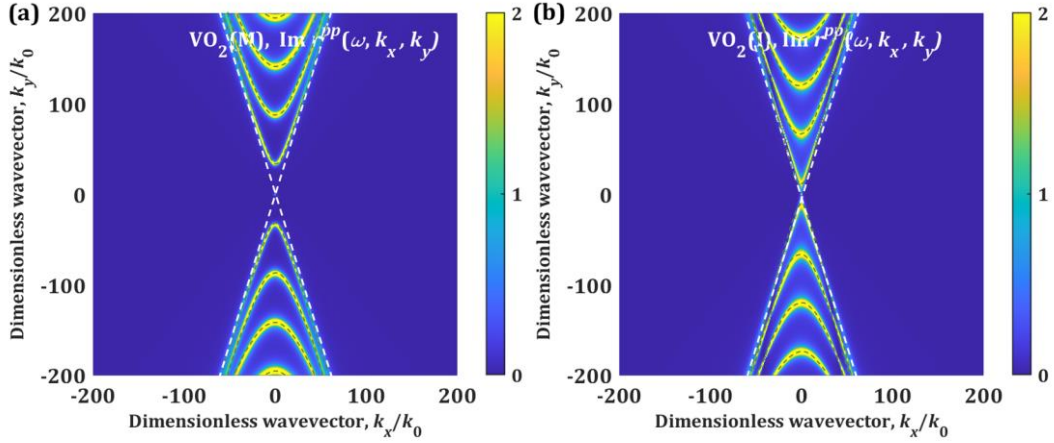

Fig. 3. Hybridization dispersions of the (a) $\alpha$-MoO$_3$/VO$_2$(M) and (b) $\alpha$-MoO$_3$/VO$_2$(I) structures varying with wavevector components $k_x$ and $k_y$ at the wavenumber of 780 cm$^{-1}$.

### 3.2 Coupling hybridization dispersion between $\alpha$-MoO$_3$ PhPs and graphene SPPs

The previous study clarified active tunability of $\alpha$-MoO$_3$ PhPs can be achieved through the phase transition of VO$_2$ without changing the structural parameters. Since graphene supports SPPs in the mid-infrared and terahertz region [18, 19], which provides a flexible way to investigate coupling hybridization dispersion modes in $\alpha$-MoO$_3$, two heterostructures are proposed to realize coupling hybridization dispersion between $\alpha$-MoO$_3$ PhPs and graphene SPPs. As a benchmark, a graphene layer is added on the top of the $\alpha$-MoO$_3$/VO$_2$ structure shown in Fig. 4(g) and its hybridization dispersion modes will be investigated. Here, $\mu_g$ = 0.3 is chosen and fixed to investigate coupling hybridization dispersion modes, it is worth noting that using other graphene chemical potentials can also obtain similar conclusions, and results are not shown here. At $\phi$ = 0° shown in Figs. 4(a) and 4(d), coupling hybridization hyperbolic plasmon-phonon polaritons (HPPPs) dispersion modes between $\alpha$-MoO$_3$ PhPs and graphene SPPs occur within its hyperbolic bands [41], which are not pure PhPs displayed in Figs. 2(a) and 2(d) of the $\alpha$-MoO$_3$/VO$_2$ structure. Especially within RB-2, the coupling hybridization polariton branches in the $\alpha$-MoO$_3$ are broader and move closer to the light cone because of the inherently lossy of the graphene SPPs, and the group velocity obviously increases when compared with that of the $\alpha$-MoO$_3$/VO$_2$ structure in Figs. 2(a) and 2(d). Compared to the hybridization dispersion in the $\alpha$-MoO$_3$/VO$_2$ structure, outside its RBs, surface plasmon-phonon polaritons (SPPPs) are formed by coupling with $\alpha$-MoO$_3$ PhPs through the dominant contribution of graphene SPPs, and possess positive group velocity keeping similar with graphene SPPs [41]. Similar results can also be obtained at $\phi$ = 45° (Figs. 4(b) and. 4(e)) and $\phi$ = 90° (Figs. 4(c) and 4(f)). Moreover, graphene SPPs characteristics in the coupled polariton dispersion become more obvious and move closer to the light cone with the increase of the graphene chemical potential, and results are not shown here. Besides, changing the phase transition state of VO$_2$ can also actively tune both coupled HPPPs and SPPPs dispersion modes in the graphene/$\alpha$-MoO$_3$/VO$_2$ structure without changing the structural parameters.

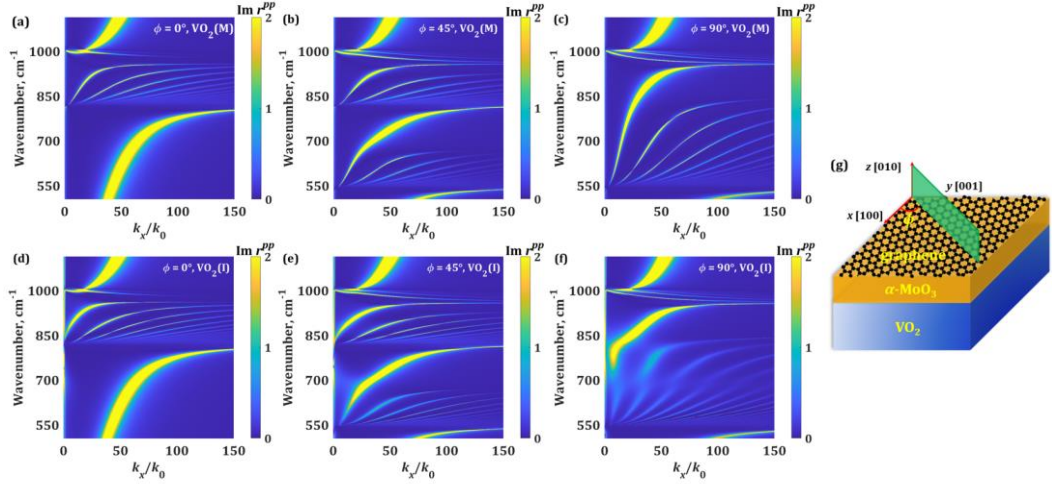

Fig. 4. Hybridization dispersion of the graphene/$\alpha$-MoO$_3$/VO$_2$ structure at (a)-(c) metal- and (d)-(f) insulator-state of VO$_2$ for different propagation plane angles $\phi$. (g) Schematic of illustration of the graphene/$\alpha$-MoO$_3$/VO$_2$ structure.

Based on the graphene/$\alpha$-MoO$_3$/VO$_2$ structure, an additional heterostructure with the placement of a surface plasmon source on either side of the $\alpha$-MoO$_3$ is considered to further investigate both HPPPs and SPPPs dispersion modes, as schematically shown in Fig. 5(g). Figs. 5(a)-5(f) show the hybridization dispersion modes of the graphene/$\alpha$-MoO$_3$/graphene/VO$_2$ structure, here keeping $\mu_{g1} = \mu_{g2} = 0.3$ is equal for simplicity. Compared with that in graphene/$\alpha$-MoO$_3$/VO$_2$ structure, the coupling effect is even more pronounced via the extra contribution of the graphene SPPs at the bottom of the $\alpha$-MoO$_3$ layer. When compared with that of the VO$_2$ substrate in insulator-state shown in Figs. 5(a)-5(c), the hybridization dispersion modes of the graphene/$\alpha$-MoO$_3$/graphene/VO$_2$(M) structure shown in Figs. 5(d)-5(f) obviously move closer to the light cone and broaden, especially for $\phi$ = 90° displayed in Fig. 5(f), which can be attributed to the reflection reinforcement at VO$_2$ layer once it changes from metal-to-insulator state [15].

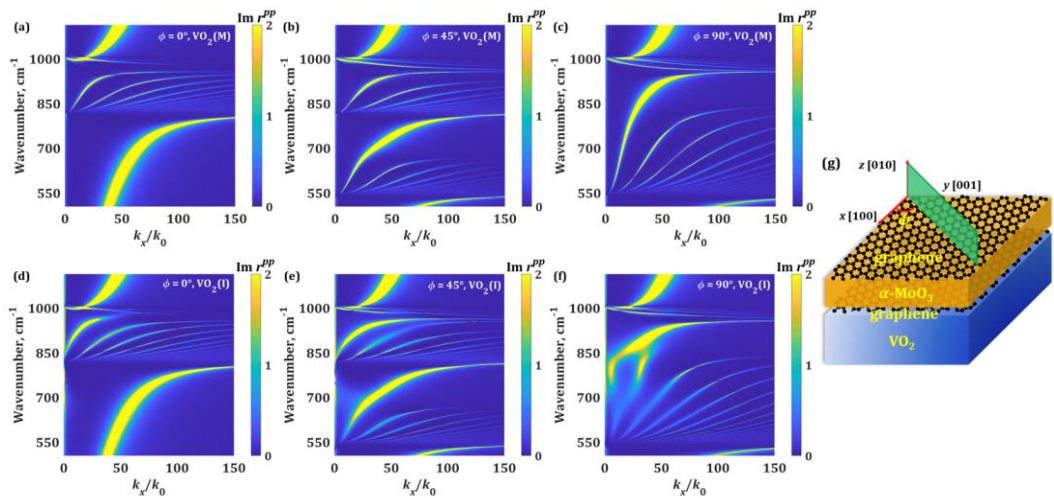

Fig. 5. Hybridization dispersion of the graphene/$\alpha$-MoO$_3$/graphene/VO$_2$ structure at (a)-(c) metal- and (d)-(f) insulator-state of VO$_2$ for different propagation plane angles $\phi$. (g) Schematic of the graphene/$\alpha$-MoO$_3$/graphene/VO$_2$ structure.

Furthermore, the hybridization dispersion modes of the graphene/$\alpha$-MoO$_3$/VO$_2$ and graphene/$\alpha$-MoO$_3$/graphene/VO$_2$ structures varying with the wavevector components $k_x$ and $k_y$ at the wavenumber 780 cm$^{-1}$ are also calculated in Figs. 6(a)-6(d) to illustrate the differences. Coupling hybridization HPPPs dispersion modes occur within the asymptotes $k_y \approx +3.26k_x$ and $k_y \approx -3.26k_x$, and coupling hybridization SPPPs dispersion modes appearing on the left and right sides of the origin is mainly derived from the contribution of graphene SPPs. When VO$_2$ in the graphene/$\alpha$-MoO$_3$/VO$_2$ structure experiences the metal-to-insulator phase transition, its hybridization HPPPs dispersion branches on the top and bottom sides of the origin gradually move towards the direction of zero-wave vector. Furthermore, as another graphene layer is added at the bottom of $\alpha$-MoO$_3$ layer in the graphene/$\alpha$-MoO$_3$/VO$_2$ structure shown in Figs. 6(c)-6(d), coupling hybridization dispersion modes between $\alpha$-MoO$_3$ PhPs and graphene SPPs are stronger, and hybridization HPPPs dispersion branches obviously move towards to the origin.

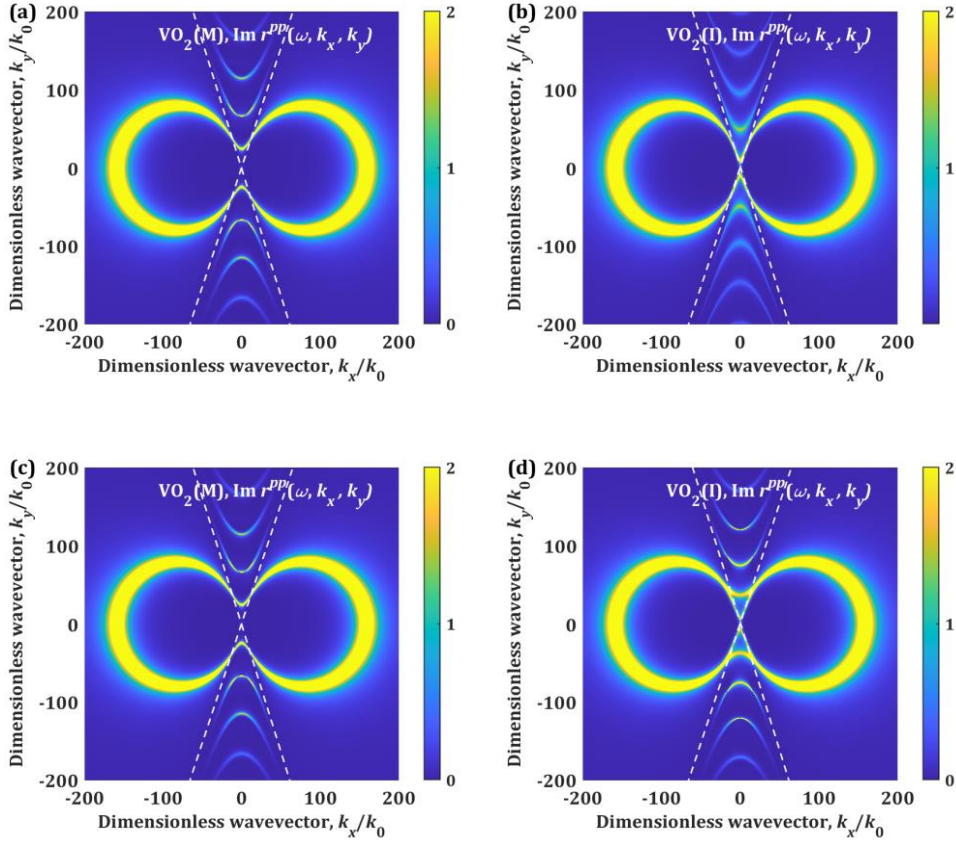

Fig. 6. Hybridization dispersions of the (a, b) graphene/$\alpha$-MoO$_3$/VO$_2$ and (c, d) graphene/$\alpha$-MoO$_3$/graphene/VO$_2$ structures varying with wavevector components $k_x$ and $k_y$ at the wavenumber of 780 cm$^{-1}$.

**3.3 Actively tuning anisotropic SE enhancement in $\alpha$-MoO$_3$-based structures**

In this section, comparative analysis of anisotropic SE enhancement in $\alpha$-MoO$_3$-based structures will be investigated to illustrate the VO$_2$ phase change can realize actively tunable SE. As compared with that in free space, the lifetime of an excited quantum emitter being placed near a special photonic material system can be modified, which is called Purcell effect

[42]. Hybridization dispersion modes modulated interaction with graphene and α-MoO$_3$ on VO$_2$ substrate has significant effects on the SE rate of an excited quantum emitter. When the moment of the dipole on a system is along the *z*-direction, its SE rate (Purcell spectrum) can be described as [43]:

$$\frac{\Gamma}{\Gamma_0} = 1 + \frac{3}{2k_0^3} \text{Re}\left(\int_0^\infty \frac{r^{pp}(\omega, k_x) e^{2ik_z d_s} k_x^3 dk_x}{k_z}\right), \quad (6)$$

here $\Gamma$ and $\Gamma_0 = \omega_0^3 |p_z|^2 / 3\pi\varepsilon_0 \hbar c^3$ are the SE rate of a dipole emitter near the system and in free space, $d_s$ is the distance between the dipole emitter and the proposed systems, $r^{pp}(\omega, k_x)$ is the total Fresnel reflection coefficient of the proposed systems [37, 38].

Then, the SE rates of α-MoO$_3$-based structures will be investigated and discussed, the distance between the dipole emitter and the proposed systems is assumed as $d_s$ = 100 nm. Firstly, the SE rates of the *α*-MoO$_3$/VO$_2$ structure shown in Fig. 1(a) are calculated at the different propagation plane angle *ϕ*, as shown in Figs. 7(a)-7(c). Results indicate SE enhancements of the *α*-MoO$_3$/VO$_2$ structure mainly gather within its hyperbolic bands, the SE enhancements of the *α*-MoO$_3$/VO$_2$ structure greatly rely on propagation plane angle *ϕ*, and its enhancement can be attributed to the contribution of *α*-MoO$_3$ PhPs, and its enhanced wavenumber of SE rate are keeping with the PhPs dispersion modes of the α-MoO$_3$/VO$_2$ structure showing in Figs. 2(a)-2(f). Moreover, the SE rate of the α-MoO$_3$/VO$_2$ structure significantly change as the VO$_2$ substrate undergoes the metal-to-insulator phase transition. For example, at *ϕ* = 0°, the maximum Γ/Γ$_0$ is 4438.8 at 980.6 cm$^{-1}$ for the α-MoO$_3$/VO$_2$(M) structure, while the maximum Γ/Γ$_0$ is 3367.2 at 984.1 cm$^{-1}$ for the α-MoO$_3$/VO$_2$(I) structure, this characteristic provides a feasible way to tune the SE rates of the *α*-MoO$_3$ actively.

Since the conductivity of graphene can be modulated by applying an extra gate voltage, which is an advantage for using graphene for photonics applications. Subsequently, the effect of the different propagation plane angle *ϕ* on SE rates of the graphene/*α*-MoO$_3$/VO$_2$ structure displayed in Fig. 4(g) are theoretically investigated, as shown in Figs. 7(d)-7(f). The SE rates of the graphene/*α*-MoO$_3$/VO$_2$ structure keep similar characteristics with that in *α*-MoO$_3$/VO$_2$ structure. Differently, a larger SE enhancement is obtained via the coupling between *α*-MoO$_3$ PhPs and graphene SPPs within its hyperbolic bands. Furthermore, when compared with that in the *α*-MoO$_3$/VO$_2$ structure, the SE rates of the graphene/*α*-MoO$_3$/VO$_2$ structure obviously strengthen outside its hyperbolic bands, which can be attributed to the contribution of HPPPs shown in Figs. 4(a)-4(f). In addition, the SE rates of the graphene/*α*-MoO$_3$/graphene/VO$_2$ structure are also considered, as shown in Figs. 7(h)-7(i). Except for the magnitude, its SE rates are basically same with that in the graphene/*α*-MoO$_3$/VO$_2$ structure, thus the graphene layer at the bottom of *α*-MoO$_3$ layer in the graphene/*α*-MoO$_3$/VO$_2$ structure just plays a role in adding flexibility. It is worth noting that increasing the thickness *t* of *α*-MoO$_3$ layer and chemical potential $\mu_g$ of graphene layer or decreasing $d_s$ can strengthen the SE rates of the proposed heterostructure, these parameters' tuning is not taken into account in this work [16].

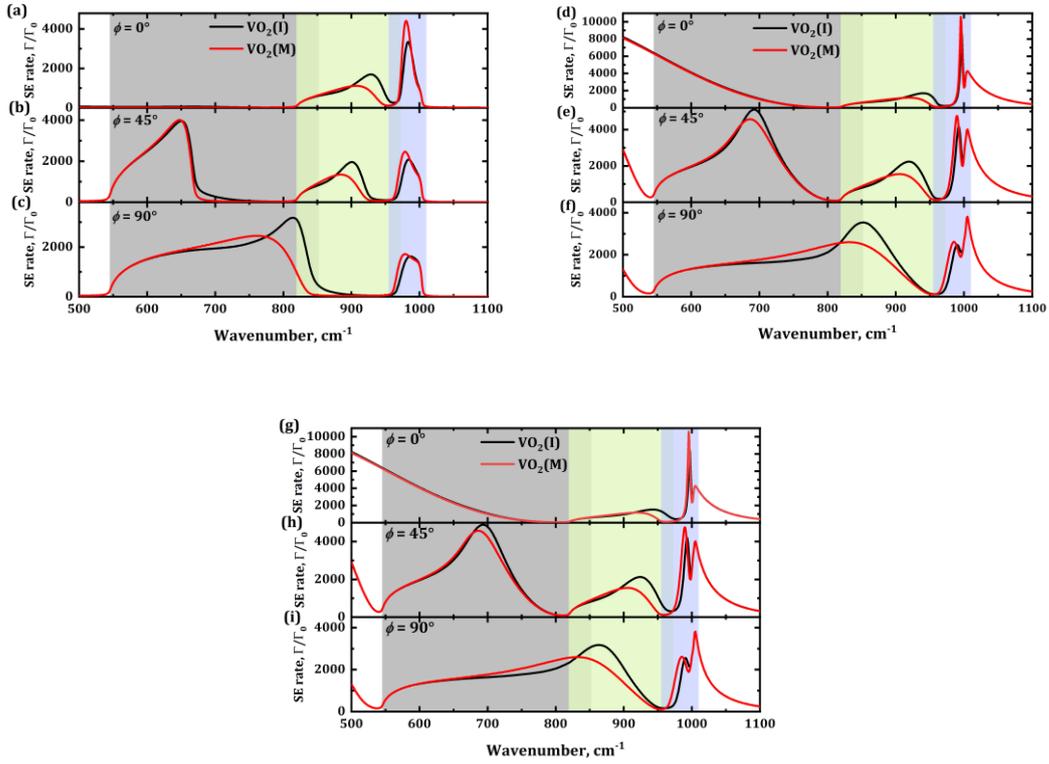

Fig. 7. SE rate of α-MoO$_3$-based structures at $d_s$ = 100 nm. (a)-(c) α-MoO$_3$/VO$_2$ structure, (d)-(f) graphene/α-MoO$_3$/VO$_2$ structure, (h)-(i) graphene/α-MoO$_3$/graphene/VO$_2$ structure.

### IV. Conclusion

In summary, a theoretical platform is constructed to actively tune anisotropic light-matter interaction in biaxial HM α-MoO$_3$ using phase change material VO$_2$ and graphene. For the α-MoO$_3$/VO$_2$ structure, the hybridization PhPs dispersion modes in α-MoO$_3$ exhibit obvious in-plane anisotropic behaviors, and the hybridization PhPs in α-MoO$_3$ are greatly dependent on propagation plane angle, its hybridization PhPs dispersion modes are gathering inside RB-2 and RB-3 at $\phi$ = 0° and RB-1 and RB-3 at $\phi$ = 90°, while its hybridization PhPs dispersion modes simultaneously occur in all hyperbolic bands and just span the part frequency range of RBs at $\phi$ = 45°. Meanwhile, the metal-to-insulator phase transition of VO$_2$ has a significant effect on the hybridization PhPs of α-MoO$_3$/VO$_2$ structure, and allows to realize an active tunable α-MoO$_3$ PhPs, especially at $\phi$ = 90°. Furthermore, as different graphene layers are added at the top or bottom of α-MoO$_3$ in α-MoO$_3$/VO$_2$ structure, actively tunable HPPPs inside its RBs and SPPPs outside the RBs are realized via the coupled between α-MoO$_3$ PhPs and graphene SPPs. In addition, actively tunable SE enhancement of α-MoO$_3$ can be realized via controlling the VO$_2$ phase transition and the graphene chemical potential. The high susceptibility of PhPs of α-MoO$_3$ opens new possibilities for potential applications in combination with strongly correlated quantum materials including PCMs and graphene.

### Funding

The authors acknowledge the support of the Fundamental Research Funds for the Central Universities (2022QN1017), National Natural Science Foundation of China (52106099), the Shandong Provincial Natural Science Foundation (ZR2020LLZ004).